\documentclass[fleqn,usenatbib]{mnras}

\usepackage{newtxtext,newtxmath}

\usepackage[T1]{fontenc}

\DeclareRobustCommand{\VAN}[3]{#2}
\let\VANthebibliography\thebibliography
\def\thebibliography{\DeclareRobustCommand{\VAN}[3]{##3}\VANthebibliography}

\usepackage{graphicx}	
\usepackage{amsmath}	
\usepackage{multirow}

\title[Modelling energy flux into the corona]{How numerical treatments of the transition region modify energy flux into the solar corona}

\author[T. A. Howson \& C. Breu]{
T. A. Howson,$^{1, 2}$\thanks{E-mail: tah2@st-andrews.ac.uk}
C. Breu,$^{2}$
\\
$^{1}$School of Design and Informatics, Abertay University, Bell Street, Dundee, DD1 1HG, U.K. \\
$^{2}$School of Mathematics and Statistics, University of St Andrews, St Andrews, Fife, KY16 9SS, U.K. \\
}
\date{Accepted XXX. Received YYY; in original form ZZZ}

\pubyear{2023}

\begin{document}
\label{firstpage}
\pagerange{\pageref{firstpage}--\pageref{lastpage}}
\maketitle

\begin{abstract}
The large temperature gradients in the solar transition region present a significant challenge to large scale numerical modelling of the Sun's atmosphere. In response, a variety of techniques have been developed which modify the thermodynamics of the system. This sacrifices accuracy in the transition region in favour of accurately tracking the coronal response to heating events. Invariably, the modification leads to an artificial broadening of the transition region. Meanwhile, many contemporary models of the solar atmosphere rely on tracking energy flux from the lower atmosphere, through the transition region and into the corona. In this article, we quantify how the thermodynamic modifications affect the rate of energy injection into the corona. We consider a series of one-dimensional models of atmospheric loops with different numerical resolutions and treatments of the thermodynamics. Then, using Alfv\'en waves as a proxy, we consider how energy injection rates are modified in each case. We find that the thermodynamic treatment and the numerical resolution significantly modify Alfv\'en travel times, the eigenfrequencies and eigenmodes of the system, and the rate at which energy is injected into the corona. Alarmingly, we find that the modification of the energy flux is frequency dependent, meaning that it may be difficult to compare the effects of different velocity drivers on coronal heating if they are imposed below an under-resolved transition region, even if the sophisticated thermodynamic adaptations are implemented.

\end{abstract}




\section{Introduction} \label{sec:intro}
The solar corona is the largest and hottest layer of the Sun's atmosphere and is the subject of one of the great outstanding questions in astrophysics; the coronal heating problem. This concerns how the surprisingly high temperatures of the corona are maintained against energy losses, including thermal conduction to the cooler layers below and radiative losses into Space. The complexity of the solar atmosphere has ensured that this problem has resisted decades of sustained effort including sophisticated observational, analytical and numerical studies. It is widely accepted that the required energy is ultimately injected into the atmosphere by complex convective motions at the solar surface and a wide variety of contemporary models show how a hot corona can be sustained as a result \citep[e.g.][]{Gudiksen2005, Bingert2011, Reale2016, Kanella2018, Breu2022, Hidetaka2023, Reid2023}. However, the specific nature of the energy dissipation mechanisms remains hotly contested. Thorough reviews of contemporary research in this area are presented by \citep[e.g.][]{Reale2014, Klimchuk2015, TVD2020c, Viall2021}. 

In recent decades, significant growth in computational power has enabled the solar atmosphere to be modelled with large scale, high-resolution MHD simulations. Increasingly, these simulations include the full, gravitationally-stratified atmosphere with the different layers of the atmosphere considered within a single numerical model \citep[e.g.][]{Hansteen2015, Cheung2019, Howson2022a, Robinson2022, Chen2023, Guo2023}. As each layer is associated with distinct physical processes occurring on disparate spatial and temporal scales, incorporating the complete atmosphere within one simulation represents a significant numerical challenge. Despite this, contemporary models attempt to track the flux of energy and mass from convective layers at the base of the simulation volume, through a complex and dynamic chromosphere and into the corona, which is heated to realistic temperatures \citep[e.g. simulations produced with the MuRAM and  Bifrost codes][]{Vogler2005, Gudiksen2011}. The success of these models can then be assessed by generating synthetic emission from the simulation results for comparison with real observations. As emission from the corona is sensitive to the plasma density and temperature, the exchange of mass between the atmospheric layers and energy dissipation rates are important components of coronal heating models.

One particularly challenging aspect of solar atmospheric modelling concerns the transition region. This is a narrow layer of the solar atmosphere that sits at the interface between the relatively cool and dense chromosphere and the hot and tenuous corona. Over the transition region, the plasma temperature increases by more than two orders of magnitude, over a short distance. As a result, there are very large temperature gradients which present a considerable problem for the finite difference schemes implemented within MHD solar atmospheric codes. Under-resolving these gradients can significantly impair the accuracy of simulations, including vastly under-estimating the upflow of plasma into the corona following heating events \cite{Bradshaw2013} and artificially suppressing thermal non-equilibrium cycles \cite{Johnston2019a}. A potentially naïve solution may be to simply increase the number of grid points used by the numerical schemes such that the transition region temperature gradients remain well-resolved. However, whilst this can be a prudent strategy in 1-D codes, the computational cost in three dimensions is prohibitive. In response, several numerical techniques have been developed \citep[e.g.][]{Lionello2009, Mikic2013, Johnston2017, Johnston2019b, Johnston2021}. These are described in more detail in Section~\ref{sec:nm}, however they generally work by adapting the effects of the transition region in order to accurately model the coronal response to heating events \citep[e.g. density enhancement due to the evaporation of chromospheric and transition region plasma][]{Hirayama1974, Fisher1985, Tian2018} even with relatively coarse numerical resolution. These techniques are designed to correctly track the evolution of plasma in the corona, such that synthetic emission can be generated from heating models for direct comparison with observations \citep[e.g.][]{Antolin2016, Pontin2017, Kanella2019, Warren2020}. However, they also broaden the transition region and thus modify the flux of other forms of energy (e.g. Poynting flux) from the lower atmosphere into the corona. By modifying this energy flux, the artificial TR broadening may have unintended consequences for large scale coronal heating simulations.

As these methods are being increasingly implemented in multi-dimensional models of the solar atmosphere \citep[e.g.][]{VanDamme2020, Zhou2021, Howson2022a, Keppens2023, Li2023, Pelouze2023}, it is now essential to quantify how coronal energy injection rates are affected by these thermodynamic treatments. To this end, in this paper, we consider how simple propagating Alfv\'en waves are modified as they propagate through different simulated transition regions. By using these waves as a proxy for more complex atmospheric dynamics, we will discuss the frequency and resolution-dependent effects on energy transmission. This will allow us to estimate how mechanical energy flux is affected by transition region modifications in larger and more complex models. The remainder of the article is presented as follows: In section \ref{sec:nm}, we describe our simple models, in section~\ref{sec:res}, we describe our results and in section~\ref{sec:conc}, we discuss the implications for contemporary modelling of the fully coupled solar atmosphere.

\section{Numerical methods}
\label{sec:nm}
For the majority of the simulations conducted within this article, we used the Lare2D code \citep{Arber2001, Arber2018}. The code advances the full, non-ideal and non-linear MHD equations given in normalised form by: 
\begin{equation}\frac{\text{D}\rho}{\text{D}t} = -\rho \nabla \cdot {\bf{v}}, \end{equation}
\begin{equation} \label{eq:motion} \rho \frac{{\text{D}{\bf v}}}{{\text{D}t}} = {\bf j} \times {\bf B} - \nabla P - \rho {\bf g} + {\bf F}_{\text{visc.}}, \end{equation}
\begin{equation} \label{eq:energy} \rho \frac{{\text{D}\epsilon}}{{\text{D}t}} = - P(\nabla \cdot {\bf v}) - \nabla \cdot {\bf q} - \rho^2\Lambda(T) + \eta \lvert {\bf j}\rvert^2 + Q_{\text{visc.}} + Q_{\text{bg.}}, \end{equation}
\begin{equation}\label{eq:induction}\frac{\text{D}{\bf B}}{\text{D}t}=\left({\bf B} \cdot \nabla\right){\bf v} - \left(\nabla \cdot {\bf v} \right) {\bf B} - \nabla \times \left(\eta \nabla \times {\bf B}\right), \end{equation}
\begin{equation}\label{eq:state} P = 2 k_BnT.
\end{equation}
Here $\rho$ is the plasma density, ${\bf v}$ is the velocity, {\bf j} is the current density, {\bf B} is the magnetic field, $P$ is the gas pressure, {\bf g} is the gravitational acceleration, $\epsilon$ is the specific internal energy density, $\eta$ is the resistivity, $k_B$ is the Boltzmann constant, $n$ is the number density and $T$ is the temperature. For numerical stability, small shock viscosity terms are included, which contribute a frictional force to the equation of motion (\ref{eq:motion}) and an associated small heating term to the energy equation (\ref{eq:energy}). These are described in detail in \citep{Arber2018, Reid2020}. By testing different dissipation coefficients, we confirmed that the viscous effects are small (for these transport coefficients), such that any wave damping is effectively negligible.

The energy equation includes contributions from thermal conduction, $\nabla \cdot {\bf q}$, optically thin radiation, $\rho^2 \Lambda(T)$, and a background heating, $Q_{\text{bg.}}$. The radiative loss curve is described in detail by \citet{Klimchuk2008} and the background heating term is implemented to maintain an initial equilibirum. The magnitude of this heating term is discussed in Sect.~\ref{sec:in_cond}. The vector, ${\bf q}$, represents the heat flux and is defined according to the \citet{Braginskii1965} model for thermal conduction in a magnetised plasma. In particular, {\bf q} is given by
\begin{equation} \label{eq:cond_vec}
{\bf q} = \left({\bf k} \cdot \nabla T\right){\bf n} + \nabla \cdot \left(\frac{b^2_{\text{min}}}{B^2 + b^2_{\text{min}}}\right)\kappa \nabla T,
\end{equation}
where ${\bf n} = \bf B/(B^2 + b^2_{\text{min}})$ is parallel to the magnetic field and $k = \kappa\, {\bf n}$ with $\kappa = \kappa_0 T^{5/2}$ and $\kappa_0 = 10^{-11} \text{ Jm}^{-1} \text{ K}^{-7/2} \text{ s}^{-1}$. The constant $b_{\min}$ is defined as a small constant to avoid numerical issues at magnetic null points $({\bf B} = {\bf 0})$ and in the limit $B^2 \gg b^2_{\text{min}}$, equation~\ref{eq:cond_vec} recovers the Spitzer-H\"arm parallel conductivity, with efficient heat conduction along field lines but negligible energy transfer across them. Although, no magnetic null points occur within our simulations, for completeness, we note that if ${\bf B} \to {\bf 0}$, isotropic conduction would be recovered. 

The second spatial derivatives of the plasma temperature required for advancing the energy equation~\ref{eq:energy} are particularly problematic in the transition region, where the temperature changes very rapidly with height. As discussed in Sect.~\ref{sec:intro}, spatially under-resolving this region in coronal heating models has significant consequences for plasma evolution. In this article, we consider two numerical techniques that allow more accurate modelling of coronal plasma even at lower resolutions. These are the L09 method \citep{Lionello2009, Mikic2013} and the TRAC approach \citep{Johnston2019b, Johnston2021}. The L09 method works by modifying the thermodynamics below a fixed cut-off temperature, $T_c$. Specifically, at temperatures below $T_c$, the parallel thermal conductivity is increased:
\begin{equation}
\kappa^*_{\parallel} (T) = \begin{cases}
\kappa_0 T^{5/2}, & \,\,\, T \ge T_c,\\
\kappa_0 T_c^{5/2}, & \,\,\, T <T_c,\\
\end{cases}
\end{equation}
and the optically thin radiative losses are decreased:
\begin{equation}
\Lambda^*(T) = \begin{cases}
\Lambda(T), & \,\,\, T \ge T_c,\\
\Lambda(T)\left(\frac{T}{T_c}\right)^{5/2}, & \,\,\, T < T_c.\\
\end{cases}
\end{equation}
Here, the asterisks represent the modified expressions implemented for our simulations. This will result in some broadening of the transition region, and this effect is greater for higher cut-off temperatures.  

\begin{figure*}
	\includegraphics[width=\textwidth]{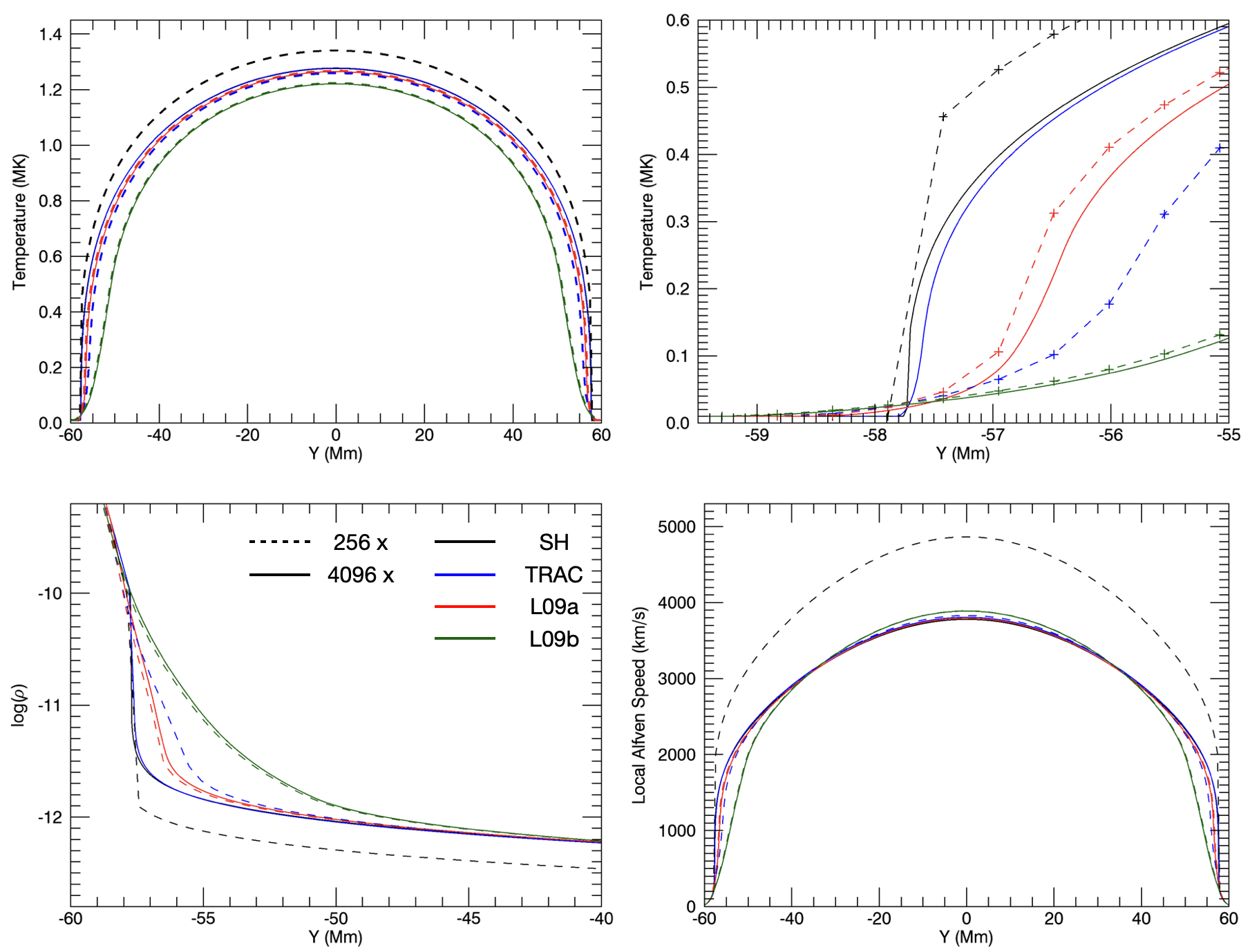}
    \caption{Initial conditions for the 256x (dashed) and 4096x (solid) resolution simulations with the SH (black), TRAC (blue), L09a (red) and L09b (green) thermodynamic treatments. The upper panels show the temperature profile over the whole loop (left) and zoomed in over the transition region (right). The crosses in the upper right hand panel reveal the simulation resolution (for the 256x case) by illustrating the location of the grid points. The lower left panel shows the logarithm of the density close to the lower foot point of the loop and the lower right panel shows the local Alfv\'en speed profile over the entire loop, for each model.}
    \label{fig:in_cond}
\end{figure*}

The TRAC method, on the other hand, uses a variable cut-off temperature $T_c$, which is continuously updated in response to simulation conditions. In particular, it sets $T_c$ as low as possible such that sufficient numerical resolution is maintained in the transition region. As such, the TRAC method results in the minimal possible transition region broadening for a given numerical resolution. Full details of the implementation are described in \citet{Johnston2019b} and \citet{Johnston2021}. We note that, in general, the TRAC approach is typically preferable as it eliminates unnecessary transition region broadening and does not require a suitable cut-off temperature to be identified a priori. In this article, we consider four different treatments for the transition region; the unmodified Spitzer-H\"arm conduction (no temperature cut-off), the TRAC treatment and two L09 cases with fixed cut-off temperatures at $T_c  = 2.5 \times 10^5$ K and $T_c = 5 \times 10^5$ K. These fixed temperature cut-offs are representative of values used in the existing literature \citep[e.g.][]{Howson2022a, VanDamme2020}. For brevity, we refer to these cases as SH, TRAC, L09a and L09b, respectively.  

In addition to these numerical treatments, thermal conduction can be limited by the maximum conductive flux that the plasma is able to support (the free-streaming limit). This occurs when the energy-transporting particles are all travelling in the same direction at the electron thermal speed. In LareXd, this saturated flux, $F_s$ is implemented as
\begin{equation}
F_s = \alpha n k_B T v_{\text{th, e}}
\end{equation}
where $\alpha$ is a user-imposed coefficient, $n$ is the number density, $k_B$ is the Boltzmann constant, $T$ is the temperature and $v_{\text{th, e}} = \sqrt{k_BT/m_e}$ is the electron thermal speed ($m_e$ is the electron mass). For the classically derived expression, where the limit is imposed by the thermal speed, a value of $\alpha = 1.5$ is required \citep[e.g.][]{Bradshaw2006}. However, it has been suggested that using $v_{\text{th, e}}/6$ is more appropriate (e.g. $\alpha = 1.5/6 = 0.25$). In either case, for the relatively cool loops considered here, the maximum conductive flux is well below the saturation threshold and thus the choice of $\alpha$ (between these two values) is moot. The choice would become significant, however, for much hotter plasma (e.g. during solar flares). For the current study, we selected the classical value, $\alpha = 1.5$.

\subsection{Initial conditions and numerical relaxation}\label{sec:in_cond}
We followed standard techniques for modelling closed, curved coronal loops as straight magnetic field lines embedded in the solar photosphere at both magnetic foot points. We consider a field line of length $l = 120$ Mm (-60 Mm $\le y \le y$ 60 Mm) which is assumed to be aligned with the central axis of a coronal loop. The magnetic field is uniform, parallel with the $y$ axis and has a field strength of 25 G. For simplicity, we neglect any expansion of the loop cross-section with height. We have implemented a range of numerical resolutions: $\{256, 512, 1024, 2048, 4096\}$ grid points along the field line. We note that although we are only considering one-dimensional problems, for consistency with existing multi-dimensional models of the Sun's atmosphere, we elected to use a two-dimensional code. However, in this paper we assume perfect invariance across the magnetic field. The cross-field direction ($x$ axis) is simulated with a nominal number of grid points (4) and periodic boundaries to maintain this invariance. The $z$ direction is perfectly invariant but the magnetic and velocity fields are permitted to have non-zero components in this direction. 

In order to mimic the curved geometry of a coronal loop, we define gravity to act in the $y$ direction with a sinusoidal profile:
\begin{equation}
g_{\parallel} = g_0 \sin\left(\frac{\pi y}{2 y_{\text{max}}}\right) ,
\end{equation}
where $g_0 \approx 274 \text{ m s}^{-2}$ is the gravitational acceleration at the solar surface ($y = \pm 60$ Mm). The acceleration points vertically downwards for $y<0$ and upwards (towards y=$y_{\text{max}}$) for $y>0$ Mm. This field-aligned component of the gravity assumes that the loop has a semi-circular profile. We neglect any cross-field contribution due to gravity. 

Coronal loops are embedded in the relatively cool and dense lower layers of the atmosphere (photosphere and chromosphere) and extend into the hot and tenuous corona. To include this in our initial conditions, we assume a temperature profile, $T = T(y)$ defined by:
\begin{equation} \label{eq:temp}
T(y) = T_{\text{ch}}  + \frac{T_{\text{co}}  -  T_{\text{ch}}}{2} \left\{\tanh \left(\frac{y + a}{b}\right) - \tanh\left(\frac{y-a}{b}\right)\right\},
\end{equation}
where $T_{\text{ch}} = 2 \times 10^4$ K is the chromospheric temperature, $T_{\text{co}} = 10^6$ K is the initial coronal temperature, $a = 55$ Mm controls the location of the transition region and $b = 0.5$ Mm controls the initial width of the temperature transition. In \cite{Bradshaw2013}, the authors detail the numerical resolution requirements for accurately reproducing coronal evolution in gravitationally stratified one-dimensional solar atmospheric loops. As discussed in Sect.~\ref{sec:intro}, spatially under-resolving the transition region will lead to lower evaporative upflows of dense plasma in response to heating, and ultimately, lower coronal densities. According to Table 1 in \cite{Bradshaw2013}, for coronal loops with relatively cool apex temperatures (3 MK), a grid resolution of approximately 25 km in the transition region is sufficient. Given that the loops in our current study are cooler (1 MK), the resolution requirements are less stringent. As our highest resolution case has a grid size of $120 \text{ Mm}/4096 \approx 29$ km, it is likely able to reproduce coronal behaviour accurately. For the lower resolutions cases, however, accurate coronal apex densities and temperatures can only be reproduced with the modifications in thermodynamics discussed above.

To find the loop's initial density profile, we assumed hydrostatic force balance and solved the resulting differential equation using a Runga-Kutte scheme. Whilst this ensures the system is in pressure balance, the conduction, radiation and heating terms in the energy equation \ref{eq:energy} mean it is not in thermodynamic balance. In order to achieve this for our initial equilibrium we then perform a numerical relaxation using the Lare2d code. In this article we set the background heating to be $Q_{\text{bg.}} =  5 \times 10^{-6} \text{ J m}^{-3} \text{ s}^{-1}$ in all cases, which generates apex temperatures of approximately 1.3 MK (see upper left panel of Fig.~\ref{fig:in_cond}). We impose a large viscosity term to damp field-aligned flows until the system settles into a state of numerical equilibrium and we take this as the initial conditions for our simulations. This relaxation viscosity is then set to zero for the subsequent wave driving simulations (Sects.~\ref{sec:driven_wave} \& ~\ref{sec:broadband}). 

The resulting density and temperature profiles are shown in Fig.~\ref{fig:in_cond} for the 256x (dashed lines) and 4096x (solid lines) resolution cases with each of the thermodynamic treatments. The upper left panel shows the temperature profiles along the full loop length. We note that the majority of the curves are very similar, closely approximating the profile of the high resolution SH model (solid black line). The exceptions are the 256x SH case (dashed black line) and the two L09 simulations (green lines). The 256x SH case significantly underestimates the coronal density \citep[see lower left panel and][]{Bradshaw2013} and, as a result, coronal radiative losses will be much lower in this case (the losses scale with the square of the density, see equation~\ref{eq:energy}). Even though energy losses in the corona are dominated by conduction, the reduction in the radiative cooling rate is sufficient to allow noticeably higher temperatures in this setup. Conversely, the thermodynamic treatment in the L09b simulations leads to significant over-broadening of the transition region (see green curves in right hand panel). This ultimately leads to the lower coronal temperatures, demonstrating that cut-off temperatures which are too high can be unhelpful for coronal loop models. However, such high cut-offs may remain necessary for very high temperature loops where the resolution requirements are even more demanding.

The effects of each treatment on the transition region are clear in the zoomed-in temperature profile displayed in the upper right panel of Fig.~\ref{fig:in_cond}. At both resolutions (256x and 4096x) the SH treatments produce the steepest transition regions with broadening evident for all other cases. The broadening is almost independent of resolution for the fixed temperature cut-off cases (L09a in red and L09b in green), however, we see much more broadening in the low resolution TRAC case (dashed blue) than in the high resolution equivalent (solid blue). This is because the temperature cut-off adjusts in the TRAC treatment according to the numerical resolution and the properties of the loop (in order to ensure coronal evolution is modelled accurately). As the SH and TRAC transition regions are very similar for the 4096x case (compare solid blue and black curves), we will assume that this represents sufficient resolution for this loop and thus use the 4096x SH case (solid black) as a benchmark simulation representing the {\it correct} evolution.

The lower left panel of Fig.~\ref{fig:in_cond} shows the logarithm of the plasma density in the bottom 20 Mm of the simulation domain. Again, this shows the effects of the thermodynamic treatment on the transition region broadening. Both SH cases (solid and dashed black lines) and the high resolution TRAC case (solid blue line) produce a steep transition region, where the density decreases by approximately two orders of magnitude over a very narrow layer. In all other simulations, this decrease occurs over a wider region, allowing for the gradients to be better resolved in the low resolution cases. We note that despite this broadening, all simulations (with the exception of the 256x SH case) exhibit similar density profiles in the corona (e.g. -40 Mm $\le y \le $ 40 Mm). These density profiles are associated with variations in the local Alfv\'en speed, $v_A = B/\sqrt{\mu_0\rho}$, displayed in the lower right hand panel of Fig.~\ref{fig:in_cond}. We note that the significantly lower coronal density attained in the 256x SH setup, produces much higher coronal Alfv\'en speeds than in all the other cases. In comparison, all other models generate similar Alfv\'en speed profiles, especially in the upper atmosphere (e.g. -40 Mm $\le y \le$ 40 Mm). However, as we discuss in detail below, even the relatively small differences that persist are sufficient to cause significantly different wave dynamics within each model.

\subsection{Boundary conditions}
We generated Alfv\'en waves within our model, by transversely perturbing the magnetic field at the lower $y$ boundary in the invariant $z$ direction. We consider wave drivers of the form 
\begin{equation} \label{eq:wave_driver}
v_z(t) = v_0 \sin \omega t,
\end{equation}
where $v_0$ is a small amplitude to generate linear waves and $\omega$ is the wave frequency. For the simulations described below, we use $v_0 = 100 \text{ m s}^{-1}$ which is less than 1\% of the minimum Alfv\'en speed in each simulation and much smaller than the coronal Alfv\'en speeds (see lower right panel of Fig.~\ref{fig:in_cond}). For the frequency, $\omega$, we consider the natural frequencies of the system and this is discussed in more detail below. At the other foot point of the loop, we impose ${\bf v} = {\bf 0}$ for the velocity and a zero-gradient condition for all other quantities. This fixes a wave node at this upper boundary. As we are using a two dimensional MHD code, we also define the $x$ boundaries to be periodic. This maintains the invariance in this direction. 

\begin{figure*}
	\includegraphics[width=\textwidth]{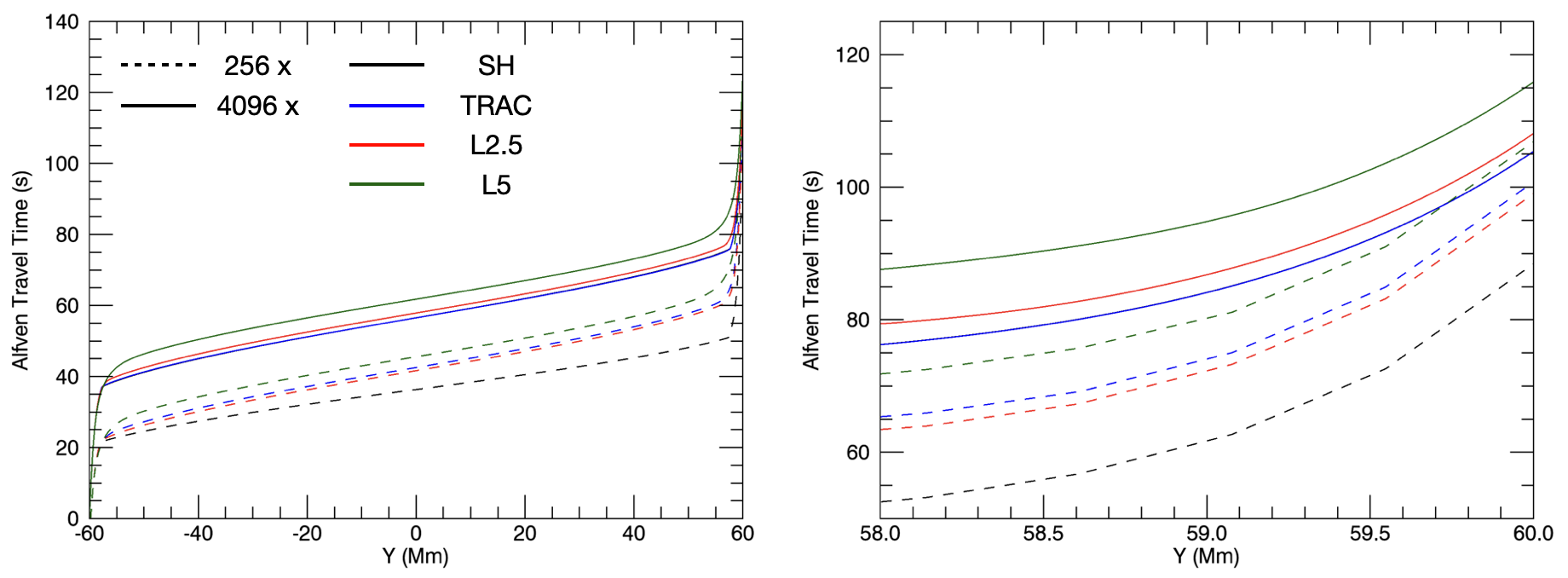}
    \caption{Alfv\'en travel time, $\tau(y)$ (see equation~\ref{eq:alf_travel_time}, for the 256x (dashed lines) and 4096x (solid lines) resolutions with the SH (black), TRAC (blue), L09a (red) and L09b (green) thermodynamic treatments. The left-hand panel shows the travel time over the whole loop and, for clarity, the right hand panel restricts the domain to the region close to the upper $y$ boundary. We note that the 4096x SH curves (solid black) is not visible as they almost exactly coincide with the 4096x TRAC curves (solid blue).}
    \label{fig:alf_travel_time}
\end{figure*}

\section{Results}
\label{sec:res}
In order to illustrate why the different transition region treatments can have a large influence on Alfv\'en wave dynamics (and hence, energy flux) in coronal loops, we first begin by considering the Alfv\'en travel time for each of our initial conditions. This represents how long it takes an Alfv\'en wave front to propagate from the driven foot point to a particular height. In other words, we calculate the travel time $\tau(y)$ as
\begin{equation} \label{eq:alf_travel_time}
\tau(y) = \int_0^y \frac{d \mathrm{s}}{v_A},
\end{equation}, 
where $v_A$ is the local Alfv\'en speed and $s$ is an infinitesimal field line element. We show the function, $\tau(y)$, for a variety of our initial conditions in Fig.~\ref{fig:alf_travel_time}. The left hand panel shows the profile over the entire length of the loop and, for clarity, the right hand panel shows $\tau(y)$ close to the non-driven foot point. We see that the total travel time varies across the simulations from approximately 88s (for the low resolution SH case) to around 116 s (for the high resolution L09b case). We note that here we have only displayed the lowest (256x) and highest (4096x) numerical resolutions and all other cases lie within the two extremes for any given thermodynamic treatment. We see that the right hand panel clearly demonstrates the significant differences between the Alfv\'en travel times for the different resolutions and transition region treatments. There are two pertinent points which drive these differences. Firstly, the different Alfv\'en speed profiles in each case (as discussed for the lower right panel of Fig.~\ref{fig:in_cond}), naturally lead to different Alfv\'en travel times. For example, the low coronal densities in the 256x SH simulation (dashed black line) result in higher Alfv\'en speeds and shorter travel times. We also note that even in cases where the coronal Alfv\'en speeds are very similar, differences in the wave speed in the transition region can result in discrepancies in the total travel time. For example, the SH and L09a treatments exhibit around a 10\% difference in the travel time despite the relatively short length of the transition region (it is much less than 10\% of the volume even in the broadened cases). Differences in Alfv\'en speeds in the lower atmosphere can have a significant impact due to the low speeds and, hence relatively long time that the wave takes to propagate through this region.

Secondly, the numerical resolution can modify the travel time simply because the local Alfv\'en speed is only known at a relatively low number of grid points. In practice, the integral in equation~\ref{eq:alf_travel_time} is calculated as a discrete summation over the simulation grid points. Therefore, in the transition region, where the wave speed changes rapidly, the low resolution simulation does not track the travel time well. As a result, the low resolution curves (dashed lines) show different travel times to their higher resolution counterparts. This effect is clearest for the fixed temperature cut-off models (as the cut-off does not change as a function of resolution), L09a (red) and L09b (green curves) which show very similar density and Alfv\'en wave profiles (see lower row of Fig.~\ref{fig:in_cond}) but still produce different travel times. Using sub-grid interpolation methods for equation~\ref{fig:alf_travel_time} would reduce this difference in the calculated Alfv\'en travel times but this would not reflect the wave propagation as simulated by the Lare2d code.

\subsection{Eigenfrequencies and eigenmodes} \label{sec:eigenfrequencies}
We can use the Alfv\'en travel times calculated above, together with the WKB approximation, to provide estimates for the natural frequencies of the coronal loops. However, these will likely be inaccurate (especially for the fundamental mode and low number harmonics \citep[e.g.][]{Wright1990} due to the significant non-uniformity of the wave speed along the modelled loops. Instead, we calculate the eigenfrequencies and corresponding eigenmodes of the field lines by considering the wave equation for non-constant propagation speed:
\begin{equation}
\frac{\partial^2 v_z}{\partial t^2} = v_A^2(y) \frac{\partial^2 v_z}{\partial y^2},
\end{equation}
where $v_A$ is the local Alfv\'en speed displayed in the lower right panel of Fig.~\ref{fig:in_cond}. Then, by considering non-trivial, oscillatory, separable solutions, $v_z = Y(y)T(t)$, of this partial differential equation, we can express the spatial variation, $Y(y)$ as
\begin{equation} \label{eq:eig_mode}
\frac{\mathrm{d}^2Y}{\mathrm{d} y^2} + \frac{\omega^2}{v^2_A(y)}Y = 0,
\end{equation}
where $\omega$ is a real constant. The Alfv\'en eigenmodes and corresponding eigenfrequencies are given by functions, $v_z(y)$, and constants, $\omega$, such that there are wave nodes ($v_z = 0$) at the two field line foot points. In order to find these, we implement a shooting method to find numerical solutions of equation~\ref{eq:eig_mode}.

\begin{figure}
	\includegraphics[width=\columnwidth]{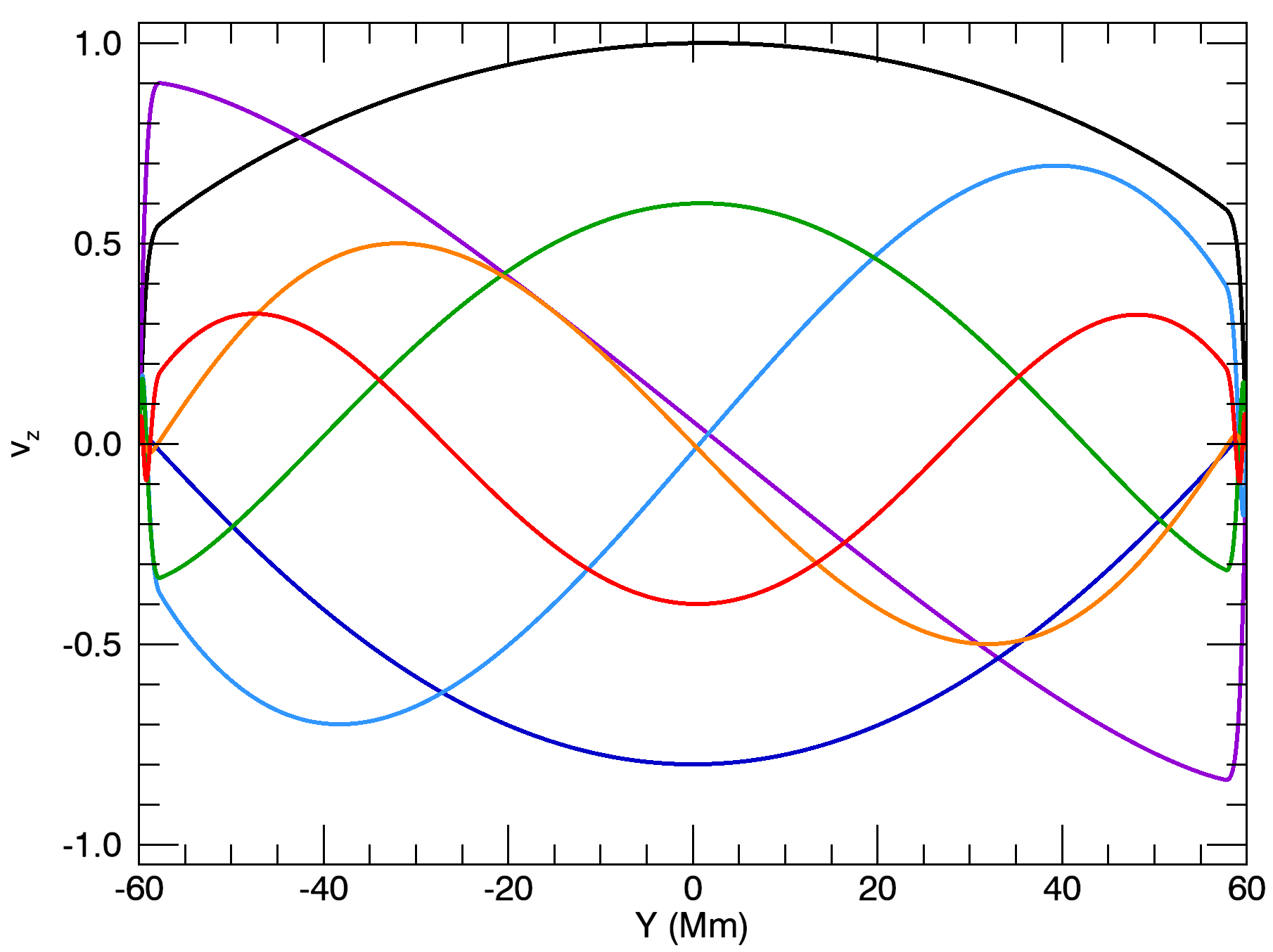}
    \caption{Eigenmodes of the fundamental (black) mode and first (purple), second (dark blue), third (light blue), fourth (green), fifth (orange) and sixth (red) overtones.These are eigenmodes of equation~\ref{eq:eig_mode} calculated for the 4096x SH model.}
    \label{fig:harmonics}
\end{figure}

In Fig.~\ref{fig:harmonics}, we display the eigenmodes of the first seven harmonics for the 4096x SH (benchmark) simulation. For clarity, we have normalised the maximum of each curve to $1 - n/10$, where $n$ is the harmonic number beginning with $n=0$ for the fundamental mode, $n=1$ for the first overtone, and so on. As such, the amplitude of each eigenmode is arbitrary. We note that due to the relatively low Alfv\'en speeds in the chromosphere and lower transition region, for higher overtones, the majority of the wave nodes are located close to the two loop foot points. As such, low resolution simulations (e.g. 256x) will not be able to resolve the short wavelengths in the chromosphere for higher frequency modes. This is discussed in more detail in Sect.~\ref{sec:broadband}. We also note that, due to the high density at these low altitudes, the magnitude of the eigenmodes here are much smaller than in the coronal volume. 

\begin{figure*}
	\includegraphics[width=0.85\textwidth]{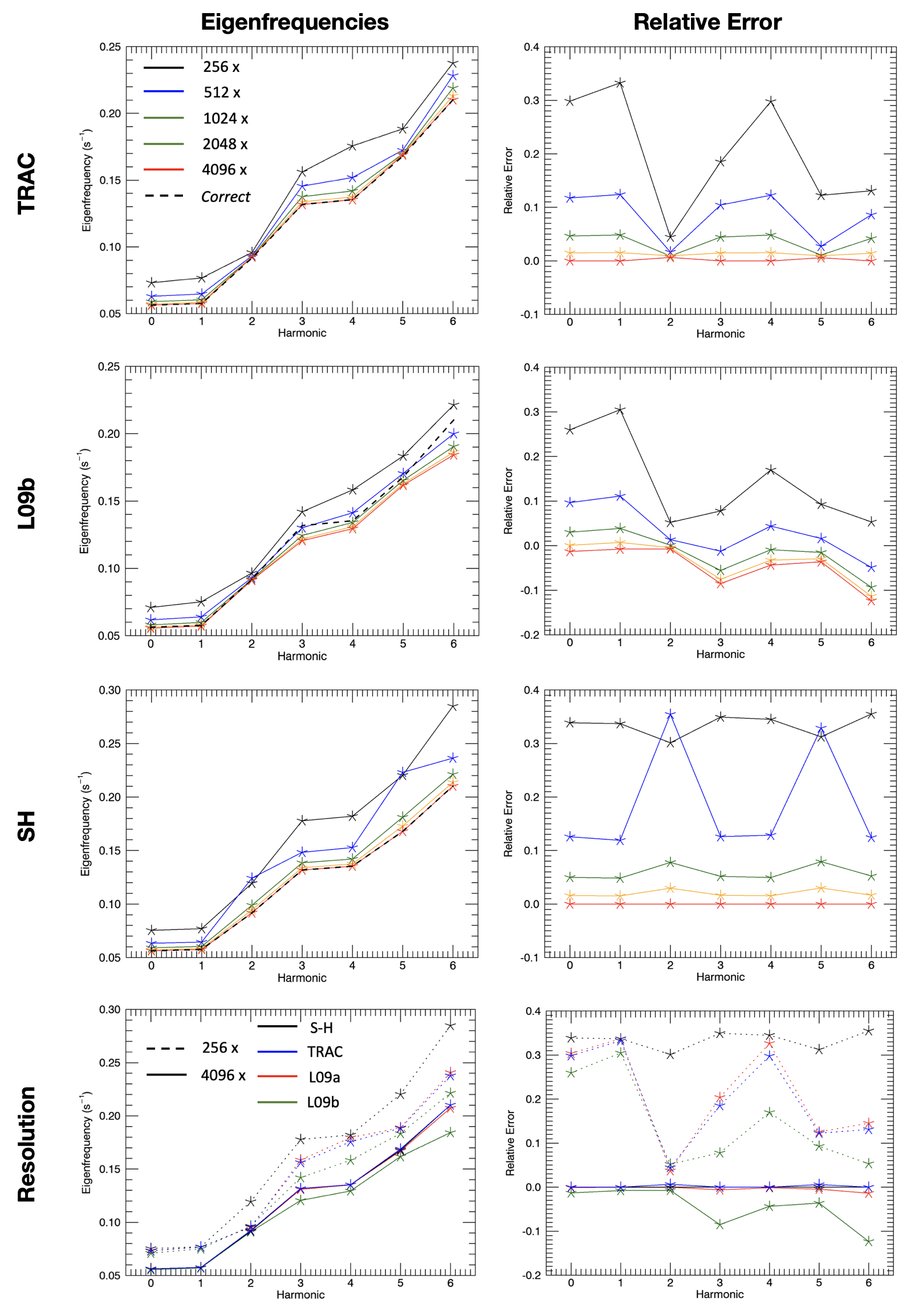}
\caption{{\emph{Left column:}} Eigenfrequencies of the fundamental (zeroth harmonic) and higher overtones for loops modelled with different numerical resolutions and thermodynamic treatments. In the upper three panels, we show the effects of resolution on loops modelled with the TRAC, L09b and SH treatments, respectively. For these panels, the dashed black line shows the 4096x SH frequencies which we use as a benchmark. The colour used for each resolution is consistent across these panels. The bottom left panel shows the effects of the thermodynamic treatment on the eigenfrequencies in the 256x (dashed lines) and 4096x (solid line) models. {\emph{Right column:}} The relative error between the curves in the adjacent panels in the left-hand column and the benchmark case (4096x SH).}
    \label{fig:harmonics_res_treatment}
\end{figure*}

We note that the exact forms of these eigenmodes will be sensitive to a wide variety of factors, including but not limited to the relative sizes of the chromosphere and corona, loop curvature, loop asymmetry and loop expansion. As such, they may not be representative of eigenmodes in real atmospheric loops. However, here we simply wish to compare the effects of the numerical resolution and thermodynamic treatment on the natural frequencies of the modelled loops. To this end, in the left hand column of Fig.~\ref{fig:harmonics_res_treatment}, we display the eigenfrequencies for the first 7 harmonics (including the fundamental mode as the zeroth harmonic) for different numerical resolutions and thermodynamic treatments. In the upper three panels, we show how these vary for different numerical resolutions in the TRAC, L09b and SH cases, respectively. In each panel, we also include the eigenfrequencies for the 4096x SH simulation as a benchmark (dashed line). For clarity, the right hand column of Fig.~\ref{fig:harmonics_res_treatment} shows the relative error between each calculated eigenfrequency and the benchmark solution. It is clear that as the resolution is increased, the TRAC (first row) and SH (third row) treatments converge to the benchmark eigenfrequencies. However, for the resolutions typically attainable in large scale, 3D MHD codes (e.g. corresponding to 256 or 512 grid points in the loop-aligned direction), the error in the frequency calculation can be approximately 40\%. At low resolutions, all thermodynamic treatments produce poor estimates of the eigenfrequency calculated for the fully resolved loop (particularly for higher harmonics). The relative error for each of these cases will depend on how important the transition region is for determining the eigenfrequency of a given loop. In particular, the relative accuracy may depend on loop lengths and temperatures and the depth of the chromosphere. We also note that for the L09b treatment (second row), whilst the eigenfrequencies do converge at high resolution, they do not converge to the benchmark case. This is easy to see at higher harmonics. This behaviour arises because the fixed temperature cut-off consistently over-broadens the transition region even at high resolutions. This phenomenon is significantly reduced (albeit still present, particularly for higher harmonics) in L09a case due to the reduced broadening of the transition region. 

In the bottom row of Fig.~\ref{fig:harmonics_res_treatment}, we display the effects of the thermodynamic treatment on the calculated eigenfrequencies (left panel) for the two extreme resolutions (256x, dashed lines; 4096x, solid lines) and the relative error (compared to the benchmark solution). At low resolution, we see that the choice of thermodynamic treatment has significant implications for the natural frequencies of the system (particularly at higher harmonics) and, as discussed before, that no choice accurately reproduces the benchmark solution. At high resolution, however, all methods (except L09b) reproduce the eigenfrequencies reasonably accurately (with small errors at high harmonics for the L09a case, due to the fixed transition region broadening). This is not a particularly surprising result; modifying the Alfv\'en speed profile in the lower atmosphere will certainly impact on the natural frequencies of the modelled loop. However, this can have important consequences for energy flux in solar atmospheric models as different resonances can be excited in each of these setups. In Sect.~\ref{sec:broadband}, we will consider whether this can lead to systematic errors for energy injection rates in general coronal heating models. However, as a brief aside, we will first consider implications of these results for seismological inversions.

\begin{figure*}
	\includegraphics[width=\textwidth]{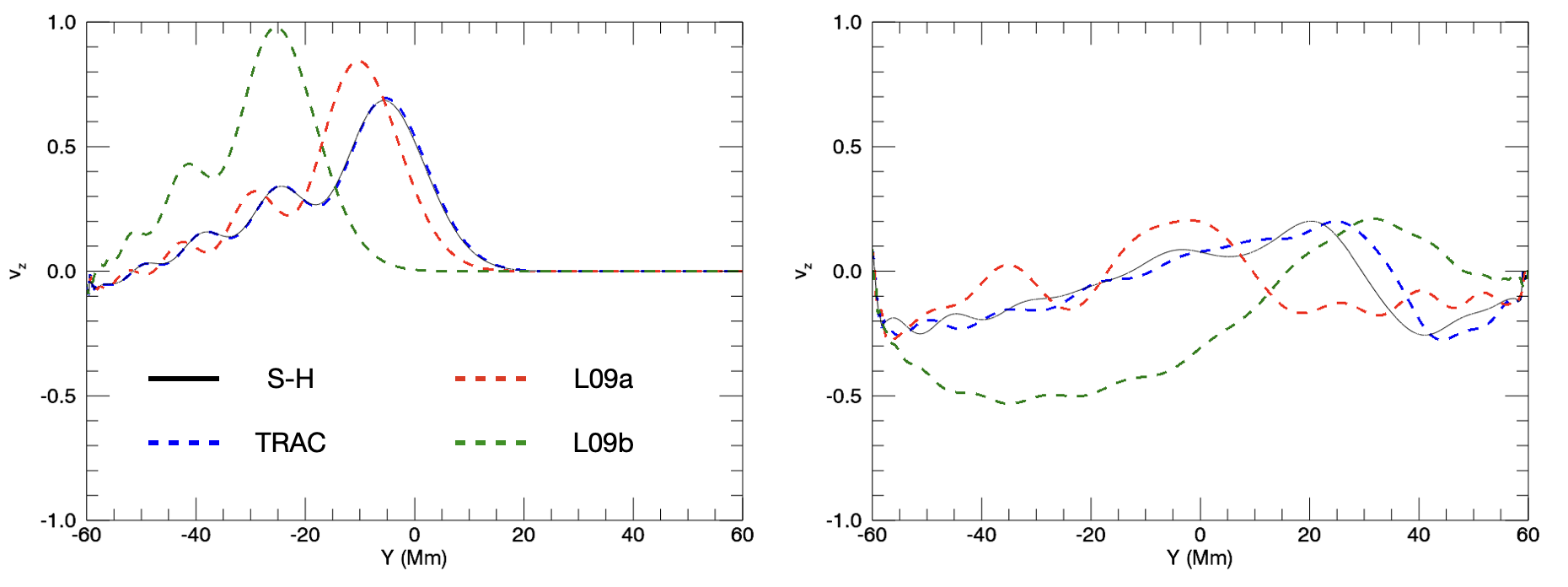}
    \caption{Transverse velocity, $v_z(y)$, excited by a continuous, high frequency, sinusoidal wave driver imposed at the lower $y$ bounndary. We show results from the 4096x resolution simulations for the SH (black), TRAC (blue), L09a (red) and L09b (green) cases. The left panel shows an early time, $t \approx 60$ s, and the right panel shows a later time, $t \approx 550$ s, after several wave reflections have occurred. The solid black line (SH) is the benchmark result.}
    \label{fig:driven_wave}
\end{figure*}

\subsection{Seismological implications}
At the broadest level, coronal seismology uses the properties of observed waves to deduce properties of the solar corona that cannot be measured directly \citep[e.g. see reviews by][]{Nakariakov2005, Andries2009, Ruderman2009, Arregui2012, DeMoortel2012, Nakariakov2020}. It is a well-developed field and has been used to provide estimates of coronal magnetic field strengths \citep[e.g.][]{Nakariakov2001, Soler2010}, the transverse density structuring of loops \citep[e.g.][]{Goddard2017, Pascoe2017, Goddard2018, Pascoe2018} and, through the use of Bayesian inference, to provide evidence for and against different solar atmospheric models \citep[e.g.][]{Arregui2011, MontesSolis2019, Arregui2022}. These studies demonstrate how expected wave behaviour (e.g. propagation speed, damping rates) derived from mathematical models can be used to deduce unknown parameters (e.g. field strength) from solar observations. However, these methods can have large uncertainties associated with them, not least because it is often difficult to definitively identify what wave mode is being observed. 

As discussed in the previous section, the lower atmosphere can play a very important role in establishing the natural frequency of these magnetic structures. They are not simply coronal loops, but are embedded in the transition region and chromosphere too. If wave nodes are established at the upper transition region, then assuming standing oscillations are purely coronal is likely a good approximation. This may be a reasonable view of oscillations excited impulsively in the corona \citep[e.g. from a nearby solar flare][]{Nakariakov1999, Li2023b}. In this paradigm, the large density gradients in the transition region may act as effective reflectors of wave energy, effectively forming wave nodes in the upper transition region. However, if these standing waves are driven by oscillatory flows in the chromosphere/photosphere \citep[as we are assuming in this article and are often invoked for driving decayless oscillations, e.g.][]{Nistico2013, Anfinogentov2015, Zhong2022, Petrova2023}, then the contribution of the lower atmosphere to the natural frequencies must be accounted for. 

As a very simple example of this idea in practice, let's suppose we observe a fundamental standing Alfv\'en wave (in reality we may be more likely to observe a kink mode but the same argument applies). For our setup, we may expect to observe a frequency of approximately $0.0563 \text{ s}^{-1}$ (the fundamental frequency for the 4096x SH model). However, these waves will typically be interpreted as a coronal-only oscillations. Thus, if we instead consider the eigenfrequencies of a coronal-only system (using equation~\ref{eq:eig_mode}), then we find a fundamental frequency of $0.0947 \text{ s}^{-1}$. Indeed this is a similar value to the frequency of the second overtone of the full system which appears to have wave nodes at the base of the corona (see dark blue line in Fig.~\ref{fig:harmonics}). As the observed frequency can be used to estimate the magnetic field strength, we will obtain an estimate that is 0.0947/0.0563 $\approx$ 1.7 times too big. This simple calculation does not consider inaccuracies in numerical wave modelling due to the thermodynamic treatments discussed throughout this article. However, despite this, the simple argument highlights an important consideration for seismological inversions; the location of the nodes for observed standing modes must be clearly identified in order to understand the true natural frequencies of the oscillating structure.   

\subsection{Modelling propagating waves} \label{sec:driven_wave}
Returning to the effects of the thermodynamic model on simulated wave dynamics, we now consider the case of a wave that is continuously driven from the lower $y$ boundary. Whereas in Sect.~\ref{sec:eigenfrequencies} we considered a purely analytical description, here we now model the propagation of Alfv\'en waves using the Lare2d code. We impose a driver as described in equation~\ref{eq:wave_driver} with frequency, $\omega = 0.069 \text{ s}^{-1}$. We note that this is a comparable to the magnitude of the fundamental frequency in each case but is non-resonant. 

In Fig.~\ref{fig:driven_wave}, we display the wave velocity ($v_z$) as a function of position along the loop for the four high resolution (4096x) simulations. The left hand panel shows the wave at an early time ($t \approx 60$ s), before the propagating wave fronts have reached the opposite foot point. The right hand panel, on the other hand, is much later, ($t \approx 550$ s), after several reflections have occurred. In both panels, we have normalised all curves by the maximum velocity obtained in the L09b simulation and the solid black line corresponds to the benchmark (4097x SH) solution. We see that at early times (left hand panel), the TRAC treatment (blue) provides very good agreement with the benchmark case. This is unsurprising given that TRAC produces minimal transition region broadening at this high resolution (compare solid blue and black lines in the upper right panel of Fig,~\ref{fig:in_cond}). The L09a (red) and L09b (green) on the other hand both show lower mean propagation speeds, with the effect greatest for the more significant broadening in the L09b case. This is in agreement with the longer travel times for these setups shown in Fig.~\ref{fig:alf_travel_time}. We also note that they both exhibit larger amplitudes, suggesting that more energy is transmitted into the corona with these treatments. We will assess this point in more detail in Sect.~\ref{sec:broadband}. The small scale oscillations that follow the leading wave front in all cases are associated with wave reflections excited as the front passes through inhomogeneities in the local Alfv\'en speed \citep[e.g.][]{Asgari-Targhi2021, Pascoe2022}.

In the right hand panel, we see that at later times that there is little agreement between with L09a, L09b and benchmark solutions. After a few reflections, the relatively small differences visible in the left panel compound to produce very different waves. Once again, the L09b case has the largest amplitude, suggesting that energy is injected into the corona more efficiently in this case. The TRAC treatment, on the other hand, still reproduces the benchmark solution with reasonable accuracy. However, there are two important caveats to note here. Firstly, as time progresses and more wave reflections take place, the differences between the TRAC and benchmark solutions will become increasingly pronounced. Furthermore, this favourable result for the TRAC method is less applicable at lower resolutions, where the transition region broadening will be significant in the TRAC case too. In general, the TRAC treatment is beneficial because it reduces broadening when possible. However, if the resolution or loop parameters are such that significant broadening is required (e.g. for high temperature loops), then the TRAC case will perform as poorly as the fixed temperature cut-offs.

\begin{figure}
	\includegraphics[width=\columnwidth]{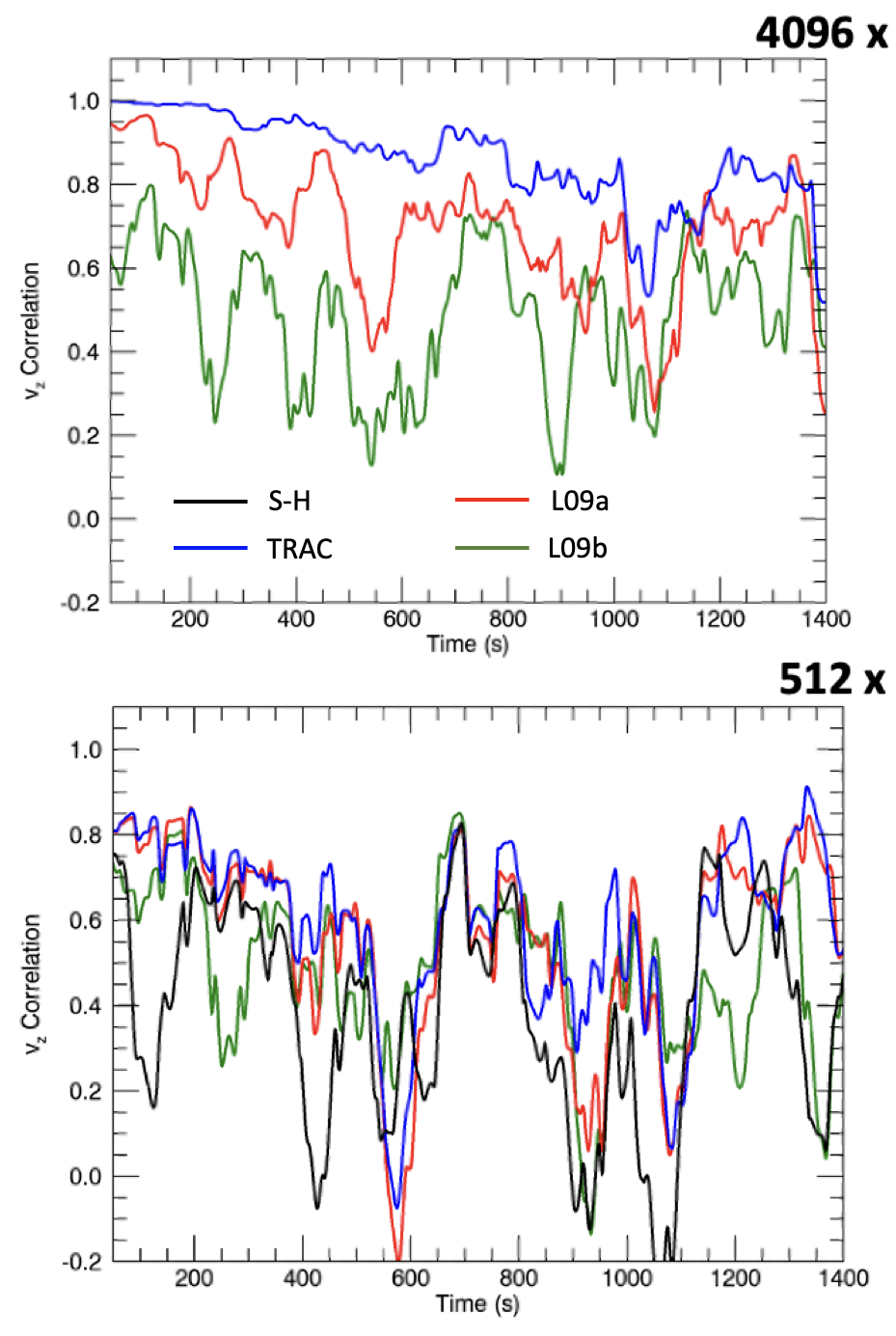}
    \caption{Pearson correlation between the wave velocity in the benchmark solution (4096x SH) and in the models with other thermodynamic treatments. The panels show how the correlation changes as a function of time at high resolution (4096x; upper panel) and at low resolution (512x; lower panel). The black curve (SH case) is not included in the upper panel because this would be measuring the correlation between one result and itself.}
    \label{fig:correlation}
\end{figure}

In Fig,~\ref{fig:correlation}, we provide a measure of the accuracy of the models produced by each thermodynamic treatment in the 4096x (upper) and 512x (lower) resolution cases. In particular, at every time output from the simulation, we measure the correlation (with the Pearson correlation coefficient) between the wave velocity ($v_z$) in each model and the benchmark solution (4096x SH). A correlation of 1 indicates a perfect match between the solutions, a correlation of 0 indicates there is no relationship between solutions and a score of -1 indicates the solutions are perfectly out-of-phase. In practice, a correlation of close to 1 indicates a good match. For the 4096x case, unsuprisingly we see that the TRAC solution with minimal transition region broadening produces a good match with the benchmark. As explained above this match steadily worsens as  time progresses. In general we see that reducing the modification of the thermodynamics is beneficial at high resolution and thus the L09a case performs better than the L09b simulation (with both worse than the TRAC case). Meanwhile, in the lower panel of Fig.~\ref{fig:correlation}, we see that for the lower resolution cases (although this still represents a high resolution in terms of large scale 3D simulations), all thermodynamic treatments produce a poor match with the benchmark solution. The reduced broadening cases (TRAC, blue and L09a, red) produce marginally better results than the SH (black) and L09b (green) cases but in general the benchmark solution is not reproduced. 

These results are concerning given that this resolution is representative of that used in state-of-the-art models of the solar atmosphere. That said, we know that many of our models do not accurately reproduce dynamics in the lower atmosphere anyway (e.g. due to neglecting partial ionisation, radiative transfer etc.). For some studies, coronal modellers may take the view that we know there is wave energy in the corona, so, as long as we model waves with the correct amplitudes/wavelengths etc., understanding precisely how waves are injected into the corona is unimportant in comparison to the dynamics (e.g. phase mixing, resonant absorption, Kelvin-Helmholtz instability) that occur as they propagate within the corona. However, increasingly coronal models are being driven by photospheric motions (e.g. through self-consistent convection or through drivers imposed below the chromosphere). Additionally, precisely tracking the energy flux through the lower atmosphere is important for understanding the energetics of the atmosphere as a whole. Indeed if the thermodynamic treatments permit an artificially high (or low) energy flux into the corona, then we will obtain incorrect heating rates, for example. There is also a more subtle issue. If the different transition region treatments are associated with errors of different magnitudes for different driver types, then comparing the effects of different drivers becomes problematic. In particular, if a low frequency driver is relatively unaffected by these errors in comparison to a high frequency driver, then making a fair comparison is challenging. In \cite{Howson2020b, Howson2022a}, the authors presented comparisons of heating associated with long (DC) and short (AC) time scale driving and found DC driving to be more effective for heating the corona. The latter study used the L09a treatment and thus it is important to establish whether the energy injection rates are artificially enhanced for the different thermodynamic treatments and whether the DC or AC driving is affected more.

\subsection{Broadband driving}\label{sec:broadband}
In order to explore this point in more detail, we will now turn our attention to a broadband wave driver to establish whether there are systematic errors in the energy injection rates. In the previous sections, we have discussed how the numerical treatment of the TR modifies the system in response to continuous transverse driving at a single frequency. However, it is unlikely that solar photospheric motions oscillate with a fixed frequency for long periods of time and thus it is important to quantify the impact of the TR treatment on energy flux for waves driven with a broadband driver. To this end, we consider drivers, $v_z(t)$ defined by
\begin{equation}
v_{z}(t) = \sum_{i=1}^N u_0 \sin\left(\omega_i t + \psi_i \right). \label{vy_eq}
\end{equation} 
These broadband wave drivers are defined as the sum of $N$ sinusoidal components with different frequencies, $\omega_i$ and phase shifts $\psi_i$. These phase shifts are randomly selected from a uniform distribution on $[0, 2\pi]$. For this article we take $N = 50$ and we restrict our consideration of the frequency space to a range between two cut-off frequencies, $\omega_{\text{min}}$ and $\omega_{\text{max}} = 3 \omega_{\text{min}}$. In particular, the $i^{\text{th}}$ frequency is defined as
\begin{equation}
\omega_i = \frac{i}{N}\left(\omega_{\text{max}}- \omega_{\text{min}}\right) + \omega_{\text{min}}.
\end{equation}

\begin{figure}
	\includegraphics[width=\columnwidth]{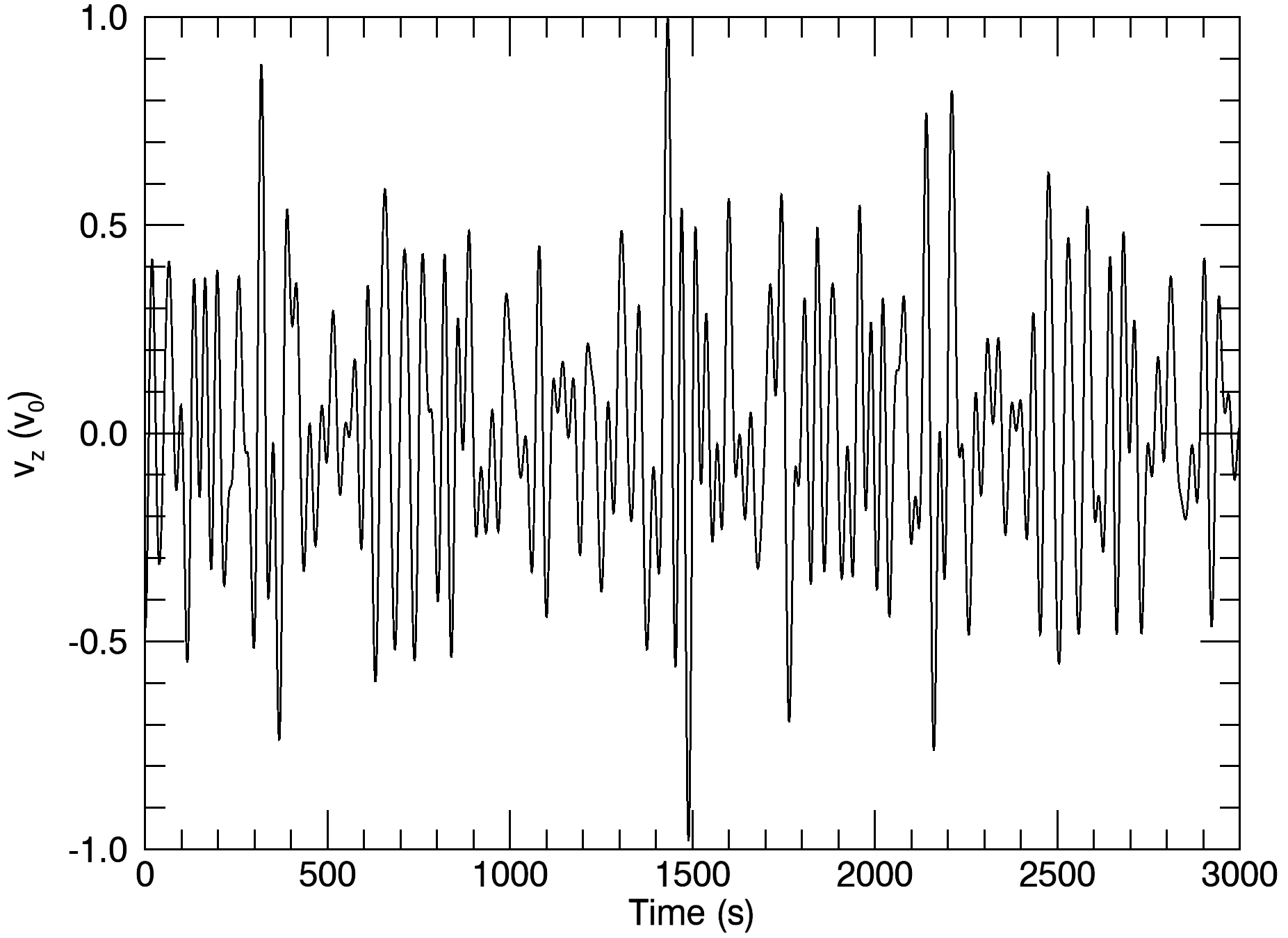}
    \caption{Temporal variation of the wave velocity imposed at the lower $y$ footpoint by the low frequency broadband driver.}
    \label{fig:bb_driver}
\end{figure}

We consider two different frequency ranges, to represent low and high frequency broadband drivers defined by $\omega_{\text{min}} = 0.8 \omega_f \approx 0.041 \text{ s}^{-1}$ and $\omega_{\text{min}} = 3.2 \omega_f \approx 0.164 \text{ s}^{-1}$. Here, $\omega_f \approx 0.0513 \text{ s}^{-1}$ is the fundamental frequency in the 4096x SH loop (see Fig.~\ref{fig:harmonics_res_treatment}). As the amplitude of each component, $u_0$, is a constant, the power in the broadband driver is independent of frequency over the range $[\omega_{\text{min}}, \omega_{\text{max}}]$. As such, within our frequency range, the wave driver has a white noise profile. The temporal profile of the driver is shown in Fig.~\ref{fig:bb_driver}.

\begin{figure*}
	\includegraphics[width=\textwidth]{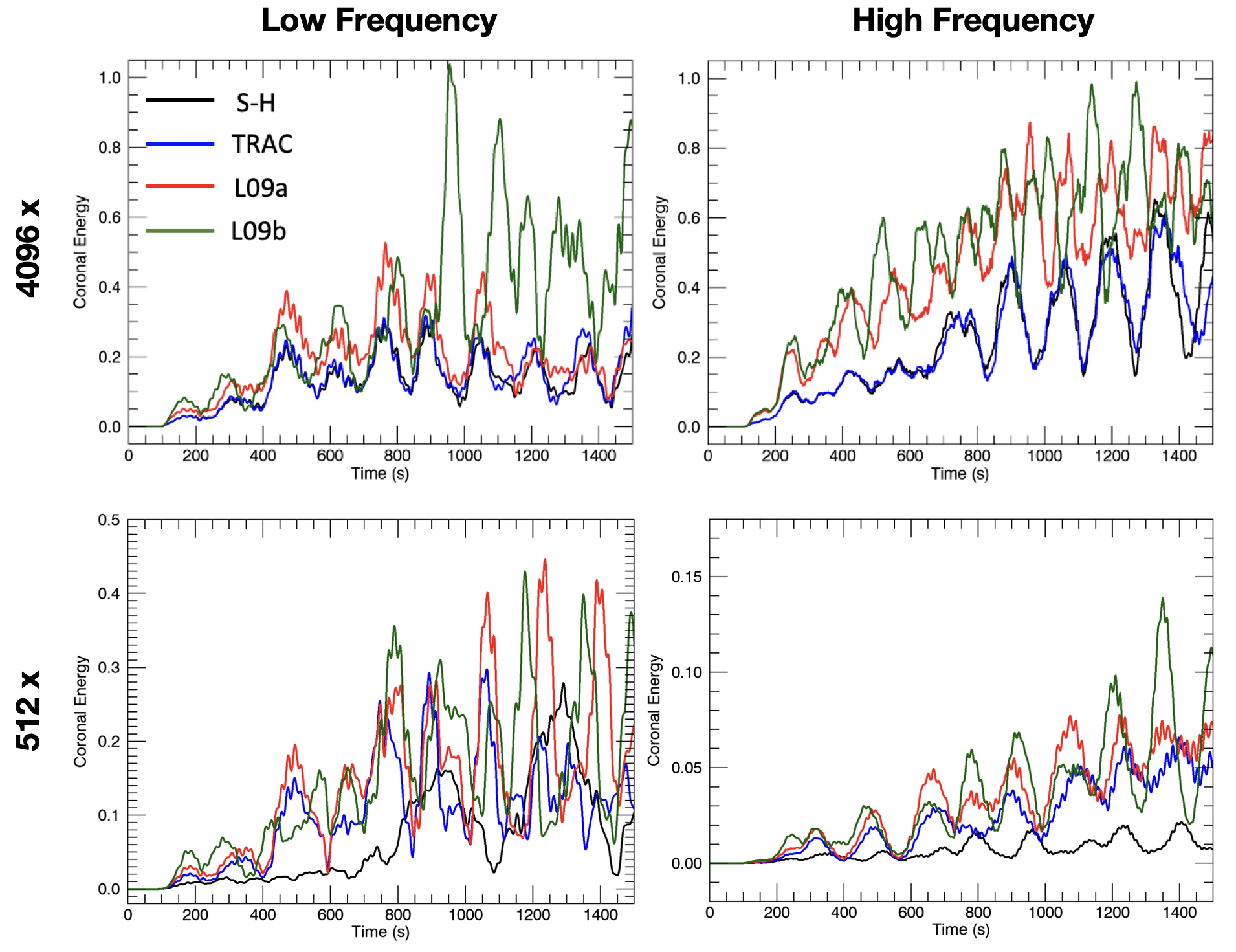}
    \caption{Energy content in the coronal portion of the domain as a function of time for low (left panels) and high (right panels) frequency drivers. We show results from the SH (black), TRAC (blue), L09a (red) and L09b (green) simulations at high (4096x, upper panels) and low (512x, lower panels) resolution. We have normalised all curves to the maximum energy content in the high frequency, high resolution L09b simulation.}
    \label{fig:Coronal_Energy_Injection}
\end{figure*}

From the discussion outlined above on Figs.~\ref{fig:driven_wave} \& ~\ref{fig:correlation}, we know that the simulated wave dynamics will be different for each thermodynamic treatment and numerical resolution. However, the important question here is whether more or less energy is injected into the corona for some treatments/resolutions in comparison to the benchmark case. In Fig.~\ref{fig:Coronal_Energy_Injection}, we show the change in the energy density, $E(t)$, within the range -40 Mm $ y $ 40 Mm as a function of time for high (4096x, upper panels) and low (512x, lower panels) resolution simulations. For this analysis we have calculated, 
\begin{equation}
E(t) = \int_{y=-40 \text{ Mm}}^{y = 40 \text{ Mm}} \left(\frac{B^2 }{2\mu_0} +  \frac{\rho v^2}{2} + \frac{P}{\gamma-1} + \rho \Phi \right)\,\,\mathrm{d} y,
\end{equation}
where the terms in the integrand are the magnetic, kinetic, internal and gravitational potential energies, respectively. For each case, we subtracted the initial value to show the change in energy density. The $y$ range for the integral ensures that we restrict our attention to the coronal subvolume of the domain. The left and right columns correspond to the low and high frequency broadband drivers, respectively. We show the results from SH (black), TRAC (blue), L09a (red) and L09b (green) simulations and all curves normalised by the maximum energy content attained in the high frequency, high resolution L09b case. 

In the high resolution, low frequency case (upper left panel), we see that the TRAC simulation (blue) reproduces the coronal energy content of the benchmark case (black curve) well. The L09a and L09b provide successively worse estimates, with the coronal energy content being overestimated by 5x at some points during the L09b case. This shows that artificially broadening the transition region permits a greater average flux of wave energy into the corona. This is because the reduced Alfv\'en speed gradients in the broadened cases reduce the efficiency of wave reflections back into the lower atmosphere. For the high frequency, high resolution cases (upper right panel), we see that the TRAC method still reproduces the benchmark solution well, however, the L09a cases now provides an equally poor estimate as the L09b simulation. The higher frequency waves have shorter wavelengths which are more liable to reflect in the transition region and hence even the weak broadening produces significant over estimations in the energy flux. This shows that the overestimation of energy injection rates can also be frequency dependent. 

In the lower resolution simulations (lower panels), we see that the mean energy injection rate is lower than in their high resolution counterparts. This effect is particularly profound for the high frequency cases (lower right panel). This is largely a consequence of relatively high numerical dissipation rates in the lower atmosphere due to the relatively low resolution and short wave lengths (particularly for the high frequency waves). This point notwithstanding, we now see different energy injection rates for all simulations and the TRAC cases no longer coincide with the SH results (or indeed with the high resolution benchmark). For both frequency drivers, we see that the wave energy in the coronal volume at any given point is very sensitive to the particular thermodynamic treatment. As with the high resolution cases, there is a tendency for broadened transition regions to permit enhanced energy flux into the corona and again these results are frequency dependent. These results show that accurately modelling MHD waves as they propagate through the lower atmosphere and into the corona is extremely challenging, particularly for high frequency modes. 

\section{Discussion and conclusions}
\label{sec:conc}
In this article, we have considered how numerical resolution and a variety of thermodynamic treatments \citep{Lionello2009, Mikic2013, Johnston2019b, Johnston2021} can modify the flux of energy from the lower solar atmosphere into the corona. As they are well understood mathematically, we have used Alfv\'en waves with a range of frequencies as a proxy for energy injection mechanisms. We have shown that the Alfv\'en travel times, eigenmodes and eigenfrequencies, and energy injection rates are all highly sensitive to the resolution and thermal conduction models used within simulations. Additionally, we have highlighted the importance of the lower atmosphere on seismological inversions when using wave modes excited beneath the transition region. 

Increasingly, numerical models of the solar atmosphere are including each of the distinct layers (photosphere, chromosphere, transition region, corona) within the simulation volume. These contemporary studies treat the physics of the lower atmosphere with varying degrees of completenness. For example, simply treating the chromosphere as a mass reservoir \citep[e.g.][]{VanDamme2020, Reid2021, Cozzo2023, Skirvin2023}, or including more accurate chromospheric physics such as radiative transfer \citep[e.g.][]{Battaglia2021, Nobrega2022, Hansteen2023, MartinezSykora2023}. Inevitably, these different approaches will lead to different energy densities at the top of the chromosphere. However, the nature of the transition region can then have further consequences for the amount of energy reaching the corona. We have shown that the choice of numerical treatment for thermodynamics in the transition region will modify the mechanical energy at higher altitudes. Furthermore, and perhaps more alarmingly, we have also shown that these treatments are associated with energy injection errors which depend on the frequency of the driver. In light of this, significant care is required when undertaking direct comparisons of coronal heating driven by different photospheric convection profiles \citep[e.g. long vs short time scale driving in][]{Howson2020b, Howson2022a}.

In the present study, we have considered loops that are relatively easy to model numerically. In particular, they are not especially hot \citep[e.g.][]{Wang2003} and they are not dynamic. As such, our negative findings may be even worse in many other situations. For example, hot loops with apex temperatures several times larger than in our models, are commonplace in active regions, and are likely important for identifying heating events \citep[e.g.][]{Bradshaw2011}. These loops require higher numerical resolution or enhanced transition region broadening (either automatically with TRAC or with higher fixed temperature cut-offs for the L09 approach), and as such will likely exacerbate our negative results. It is also important to reiterate that short wave lengths in the chromosphere (due to the relatively low Alfv\'en speeds) can lead to significant wave damping and reduced energy transmission, particularly for high frequencies. This may be unimportant for studies interested in wave propagation in the corona, however, as discussed above, this does have implications for comparing the effects of different photospheric drivers on coronal heating. 

It may seem that waves reflected at the transition region may have further opportunities to be transmitted into the corona following a second (and subsequent) reflections at the driven boundary. However, whilst this is an important and subtle point, this effect does not necessarily permit enhanced coronal energy injection over long time periods. In our simulations, upon returning to the driven boundary, reflected waves interact with the imposed velocity field. This modifies the Poynting flux injected into the domain and can have important consequences for the energetics of the system. If the reflected waves are in-phase with the wave driver, then resonances will be excited and the Poynting flux will increase. However, for non-broadband drivers this is unlikely, and the waves will typically be non-resonant. If non-resonant reflected waves have the same amplitude as the driver, then, on average, the Poynting flux will be equally positive and negative over the course of a wave period. In such a case, there will be no net energy injection after the first reflected waves reach the boundary. However, typically the reflection coefficient at the transition region will be less than unity and the reflected waves will have lower amplitudes. In this case, the driver will still inject energy into the system but at a reduced rate. As such, it may be misleading to think that the reflected waves have multiple attempts to propagate into the corona as they can also reduce the rate at which energy is injected into the system.

On the basis of our results, we recommend the use of the TRAC method as the most suitable treatment for resolutions currently attainable in large scale, multi-dimensional coronal heating models. As the TRAC approach is associated with the minimum possible broadening (for any given loop and numerical resolution), it will generally produce the smallest errors in the simulated wave dynamics. That said, at attainable numerical resolutions, it will still provide a poor match to a fully resolved benchmark case, particularly for high frequency waves. At this stage, we offer no concrete solution to these issues and believe an appropriate first step will be accurately quantifying the effects of the thermodynamic treatments for a variety of loop lengths and geometries, magnetic field strengths, wave modes (e.g. slow waves, kink waves) and also for long timescale driving (e.g. DC heating mechanisms). Only then will we be able to determine how significant these issues are for contemporary three-dimensional solar atmospheric modelling. 

\section*{Acknowledgements}
The research leading to these results has received funding from the UK Science and Technology Facilities Council (consolidated grant ST/S000402/1). The authors would like to thank Dr J Reid for his help and comments during the preparation of this manuscript. Finally, the authors would also like to thank the anonymous referee for considering our work and providing helpful suggestions to improve our article.

\section*{Data Availability}
The data from the numerical simulations and analysis presented in this paper are available from the corresponding author upon reasonable request.



\bibliographystyle{mnras}
\bibliography{transition_region_waves} 





\bsp	
\label{lastpage}
\end{document}